\documentclass[twocolumn,showpacs,aps,prl]{revtex4-1}

\usepackage{amsmath,amsfonts,amssymb}
\usepackage{wrapfig}
\usepackage{graphicx}
\usepackage{bbm}
\usepackage{color}
\usepackage{url}

\begin{document}

\title{Quantifying Quantum-Mechanical Processes}

\author{Jen-Hsiang Hsieh$^{\dag}$, Shih-Hsuan Chen$^{\dag}$, Che-Ming Li}
\thanks{cmli@mail.ncku.edu.tw $^{\dag}$These authors contributed equally to this work.}
\affiliation{Department of Engineering Science, National Cheng Kung University, Tainan 70101, Taiwan}

\begin{abstract}
The act of describing how a physical process changes a system is the basis for understanding observed phenomena. For quantum-mechanical processes in particular, the affect of processes on quantum states profoundly advances our knowledge of the natural world, from understanding counter-intuitive concepts to the development of wholly quantum-mechanical technology. Here, we show that quantum-mechanical processes can be quantified using a generic classical-process model through which any classical strategies of mimicry can be ruled out. We demonstrate the success of this formalism using fundamental processes postulated in quantum mechanics, the dynamics of open quantum systems, quantum-information processing, the fusion of entangled photon pairs, and the energy transfer in a photosynthetic pigment-protein complex. Since our framework does not depend on any specifics of the states being processed, it reveals a new class of correlations in the hierarchy between entanglement and Einstein-Podolsky-Rosen steering and paves the way for the elaboration of a generic method for quantifying physical processes.
\end{abstract}

\maketitle

\section*{Introduction}

A physical process is comprised of a series of actions that, in themselves, evolve in a way that is independent of a systems initial state. In the field of the foundations of quantum physics, there is strong interest in identifying processes that cannot be explained using classical physics. The identification of such processes helps clarify whether quantum mechanics can describe the rationale behind observed phenomena, such as transport in solid-state nanostructures \cite{Brandes05} and functional roles in biological organisms \cite{Lambert13}. Furthermore, as one wishes to take advantage of quantum-mechanical effects for some task, for instance, from atomic networks, semiconductor spintronics \cite{Coherence08}, quantum information \cite{Nielsen&Chuang00,You05,Gisin07,Ladd10,Buluta11} and quantum simulation \cite{Buluta09,Georgescu14} to the creation of  nonclassical phenomena using superconducting circuits \cite{You11,Nation12,Xiang13}, there is always a need to ensure that key procedures or processes involved in the task are reliably performed in the quantum regime. Considerable progress has been made in responding to this need \cite{Shevchenko08,Lambert10,Miranowicz10,Bartkowiak11,Miranowicz15a,Miranowicz15b}. However, characterizing the output-state responses to a process, for instance, based on imposing what can be thought of as a classical constraint \cite{Brunner14,Emary14} or through deduction from the predictions of quantum theory \cite{Guhne09,Li12}, remains a paradigm for qualitatively reflecting the existence of a nonclassical process. This approach is significant in its own right, but the most exciting aspect is the questions it raises: can a quantum-mechanical process be quantified? If so, what are the implications of such quantification?

Motivated by these questions, we present a rigorous framework for quantifying quantum-mechanical processes. This formalism simultaneously addresses a wide range of physical processes described by the general theory of quantum operations and provides benchmarks for problems of greater interest in quantum information \cite{Nielsen&Chuang00,You05,Gisin07,Ladd10,Buluta11}. It also gives insightful connections between quantum processes and the essence of other concepts, for example, non-Markovian quantum dynamics \cite{Breuer16,Vega17}. In addition, but not less importantly, this framework enables quantum states to be explored and defined to uncover new characteristics for both composite and single systems.

\begin{figure*}[t]
\includegraphics[width=17.8cm]{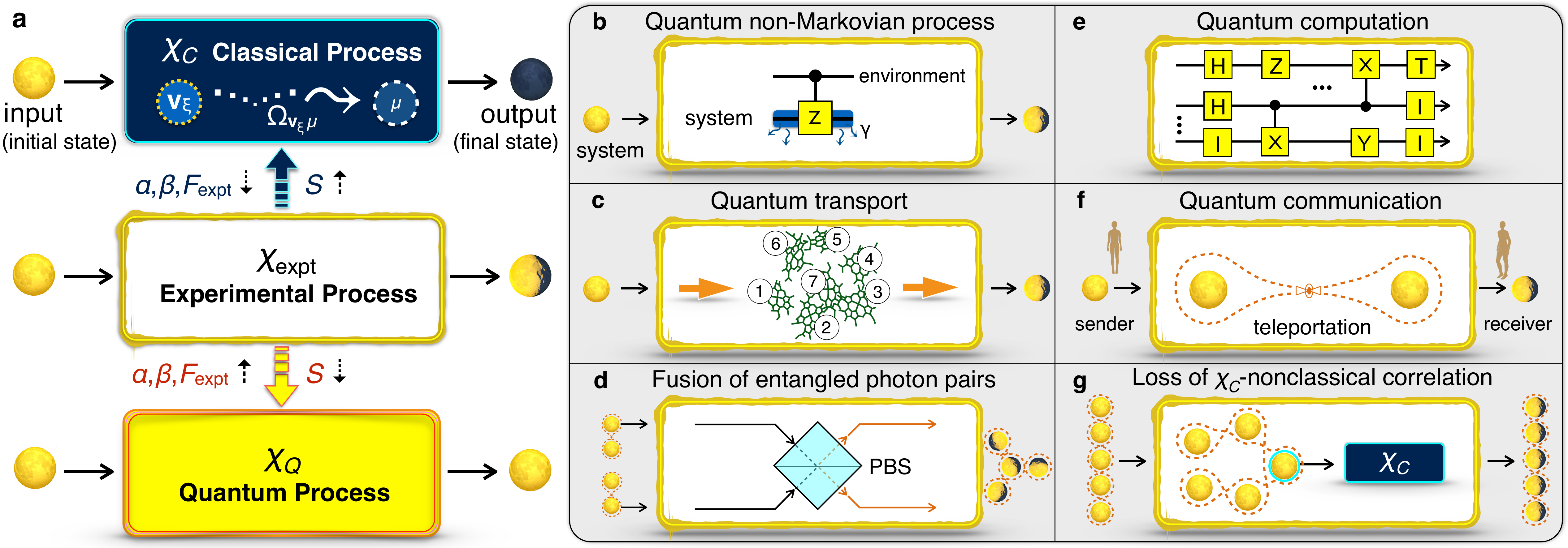}
\caption{Quantifying quantum-mechanical processes. \textbf{a} Suppose that a physical process is experimentally determined by a process matrix $\chi_{\text{expt}}$; how a system evolves from an arbitrary initial state $\rho_{\text{initial}}$ to some final state $\rho_{\text{final}}$ is specified by the process matrix $\chi_{\text{expt}}$ through the mapping $\chi_{\text{expt}}(\rho_{\text{initial}})=\rho_{\text{final}}$, which preserves the Hermiticity, trace, and positivity of the system density matrix. The amount of quantumness $\chi_{Q}$ of the process, which cannot be described at all by any classical processes $\chi_{C}$, can be characterized and quantified by $\alpha$ (composition), $\beta$ (robustness), $F_{\text{expt}}$ (process fidelity) and $S$ (von Neumann entropy). For instance, for a perfect (worst) experiment on a target quantum process, $\alpha$, $\beta$ and $F_{\text{expt}}$ will attain their individual maximum (minimum) values whereas $S$ will reach the minimum (maximum) uncertainty of the quantum process. These variables have significant applications to aid in the exploration and evaluation of all physical processes described by the quantum operations formalism, such as \textbf{b},\textbf{c} the dynamics of open quantum systems, \textbf{d} the generation of multipartite entanglement, and \textbf{e},\textbf{f} quantum-information processing. \textbf{g} This framework shows a new correlation model in the class between genuine multipartite EPR steering and genuine multipartite entanglement, called the $\chi_{C}$-nonclassical correlations.}\label{basicidea}
\end{figure*}

\section*{Results}
\subsection*{Classical processes}
We define a classical process as a set of steps involving the general descriptions of a classical state and its evolution: the initial system can be considered a physical object with properties satisfying the assumption of realism \cite{Brunner14}; then, the system evolves according to classical stochastic theory \cite{Breuer&Petruccione07} (Fig.~\ref{basicidea}a). The assumption of realism specifies that the system is in a state described by a set of measurement outcomes, for example, a set of outcomes for $N$ physical properties of interest $\textbf{v}_{\xi}\equiv(v_{1},v_{2},...,v_{N})$. If each physical property has $d$ states, then we have $d^{N}$ possible sets, $\textbf{v}_{1},...,\textbf{v}_{d^{N}}$. The dynamics of these classical states are fully described by the transition probabilities, $\Omega_{\textbf{v}_{\xi}\mu}$, from $\textbf{v}_{\xi}$ to a final state denoted by $\mu$. The above concept can be applied to the cases in which the state of the system is probabilistically prepared according to a probability distribution $P(\textbf{v}_{\xi})$. Furthermore, if we focus on a specific initial state of the $k$th physical property, e.g., $v_{k}=v'_{k}$, then the corresponding final state of the system has the form
\begin{equation}
\sum_{\mu}\sum_{\xi}P(\textbf{v}_{\xi}|v'_{k})\Omega_{\textbf{v}_{\xi}\mu}\rho_{\mu}=\sum_{\mu}\Omega_{v'_{k}\mu}\rho_{\mu},\label{resultingstate}
\end{equation}
where $\Omega_{v'_{k}\mu}=\sum_{\xi}P(\textbf{v}_{\xi}|v'_{k})\Omega_{\textbf{v}_{\xi}\mu}$. Indeed, the final states (\ref{resultingstate}) conditioned on different properties and states $v'_{k}$ can be used to capture the essence of the classical process. We use process tomography (PT), an application of the quantum operations formalism \cite{Nielsen&Chuang00}, to systematically exploit these experimentally measurable quantities and then completely characterize the classical process using a positive Hermitian matrix, called the process matrix,
\begin{equation}
\chi_{C}\big(\{\sum_{\mu}\Omega_{v'_{k}\mu}\rho_{\mu}\}_{\text{PT}}\big).\label{chic}
\end{equation}
We will hereafter use a process matrix to refer to a physical process within the text. In the following, we will illustrate the derivation of a classical process matrix.

\subsubsection*{Derivation of $\chi_{C}$ for classical processes}

In order to show explicitly how to apply PT to a classical process to completely characterize its classical features, a classical-process scenario for single two-level systems is given as a concrete example of Eq.~(\ref{chic}).
Since a classical process treats the initial system as a physical object with properties satisfying the assumption of realism, the system can be considered as a state described by a fixed set $\textbf{v}_{\xi}$. We assume that the system is described by three properties, say $V_{1}$, $V_{2}$ and $V_{3}$, where each one has two possible states. There exist $2^{3}=8$ realistic sets underlying the classical object: $\textbf{v}_{\xi}(v_{1},v_{2},v_{3})$, where $v_{1},v_{2},v_{3}\in\{+1,-1\}$ represent the possible measurement outcomes for $V_{1}$, $V_{2}$ and $V_{3}$, respectively. The subsequent classical evolution changes the system from $\textbf{v}_{\xi}$ to a final state denoted by $\mu$ according to the transition probabilities $\Omega_{\textbf{v}_{\xi}\mu}$. Such evolution can always be rephrased as the transition from a specific state set $\textbf{v}_{\xi'}$ to some final state $\mu'$ with $\Omega_{\textbf{v}_{\xi'}\mu'}=1$. Next, by using state tomography, each final state is reconstructed as a density operator $\rho_{\mu'}$. Then the states under the assumption of realism evolve according to
\begin{equation}
\begin{split}
\textbf{v}_{1}(+1,+1,+1)\rightarrow\rho_{1}&,\hspace{1pt}\textbf{v}_{2}(+1,+1,-1)\rightarrow\rho_{2},\\
\textbf{v}_{3}(+1,-1,+1)\rightarrow\rho_{3}&,\hspace{1pt}\textbf{v}_{4}(+1,-1,-1)\rightarrow\rho_{4},\\
\textbf{v}_{5}(-1,+1,+1)\rightarrow\rho_{5}&,\hspace{1pt}\textbf{v}_{6}(-1,+1,-1)\rightarrow\rho_{6},\\
\textbf{v}_{7}(-1,-1,+1)\rightarrow\rho_{7}&,\hspace{1pt}\textbf{v}_{8}(-1,-1,-1)\rightarrow \rho_{8}.
\end{split}\label{setintorho}
\end{equation}
We now consider specific states of physical properties as the input states. If we focus on a state of the third property, say $v_3=v'_3$, the final state is described as $\rho^{(c)}_{\text{final}|v'_{3}}=\sum_{\mu}\Omega_{v'_{3}\mu}\rho_{\mu}$, where $\Omega_{v'_{3}\mu}$ shows the probability of transition from $v'_3$ for all the possible sets $\textbf{v}_{\xi}$ to the final state $\rho_{\mu}$. The transition probabilities therein read $\Omega_{v'_{3}\mu}=\sum_{\xi=1,3,5,7}P(\textbf{v}_{\xi}|v'_{3})\delta_{\xi\mu}$ and $\Omega_{v'_{3}\mu}=\sum_{\xi=2,4,6,8}P(\textbf{v}_{\xi}|v'_{3})\delta_{\xi\mu}$ for $v'_{3}=+1$ and $v'_{3}=-1$, respectively. See Eq.~(\ref{resultingstate}). Since $P(v'_{3})P(\textbf{v}_{\xi}|v'_{3})=P(\text{v}_{\xi})P(v'_{3}|\text{v}_{\xi})$ and $P(v'_{3})=1/2$ under the assumption of a uniform probability distribution of $v_{k}$, the final states are written as
\begin{equation}
\rho^{(c)}_{\text{final}|v'_{3}=+1}=\!\!\!\!\!\!\sum_{\mu=1,3,5,7}\!\!\!\!\!2P(\textbf{v}_{\mu})\rho_{\mu},\rho^{(c)}_{\text{final}|v'_{3}=-1}=\!\!\!\!\!\!\sum_{\mu=2,4,6,8}\!\!\!\!\!2P(\textbf{v}_{\mu})\rho_{\mu}.\label{f3}
\end{equation}
Similarly, for the other states $v'_{1}=\pm1$, $v'_{2}=\pm1$ under the condition $P(v'_{1})=P(v'_{2})=1/2$, the classical process has the following output states:
\begin{equation}
\begin{split}
&\rho^{(c)}_{\text{final}|v'_{1}=+1}=\!\!\!\!\!\!\sum_{\mu=1,2,3,4}\!\!\!\!\!2P(\textbf{v}_{\mu})\rho_{\mu},\rho^{(c)}_{\text{final}|v'_{1}=-1}=\!\!\!\!\!\!\sum_{\mu=5,6,7,8}\!\!\!\!\!2P(\textbf{v}_{\mu})\rho_{\mu},\\
 &\rho^{(c)}_{\text{final}|v'_{2}=+1}=\!\!\!\!\!\!\sum_{\mu=1,2,5,6}\!\!\!\!\!2P(\textbf{v}_{\mu})\rho_{\mu},\rho^{(c)}_{\text{final}|v'_{2}=-1}=\!\!\!\!\!\!\sum_{\mu=3,4,7,8}\!\!\!\!\!2P(\textbf{v}_{\mu})\rho_{\mu}.\label{f12}
\end{split}
\end{equation}

The essence of PT is that a process of interest is completely characterized by a process matrix. Using the outputs of three complementary observables (e.g. the Pauli matrices $I$, $X$, $Y$, and $Z$) from the process \cite{Nielsen&Chuang00}, it is experimentally feasible to determine the process matrix. A classical process makes these observables decomposable in terms of Eqs.~(\ref{f3}) and (\ref{f12}): $I\rightarrow I_{c}\equiv\rho^{(c)}_{\text{final}|v'_{3}=+1}+\rho^{(c)}_{\text{final}|v'_{3}=-1}$, $X\rightarrow X_{c}\equiv\rho^{(c)}_{\text{final}|v'_{1}=+1}-\rho^{(c)}_{\text{final}|v'_{1}=-1}$, $Y\rightarrow Y_{c}\equiv\rho^{(c)}_{\text{final}|v'_{2}=+1}-\rho^{(c)}_{\text{final}|v'_{2}=-1}$, and $Z\rightarrow Z_{c}\equiv\rho^{(c)}_{\text{final}|v'_{3}=+1}-\rho^{(c)}_{\text{final}|v'_{3}=-1}$. Then the classical process matrix specifying how states evolve regardless of inputs can be written as the form :
\begin{equation}
\chi_C=  \left[ \begin{matrix}
    \rho_{C,00} & \rho_{C,01} \\
    \rho_{C,10} & \rho_{C,11}
    \end{matrix}
\right],
\end{equation}
where $\rho_{C,00}=(I_{c}+Z_{c})/2$, $\rho_{C,01}=(X_{c}+iY_{c})/2$, $\rho_{C,10}=(X_{c}-iY_{c})/2$ and $\rho_{C,11}=(I_{c}-Z_{c})/2$.

The above concepts and methods can be extended to multi-level and multipartite physical systems. For instance, a $d$-level system can be classically described by a fixed set $\textbf{v}_{\xi}$ with $d^2-1$ properties. As illustrated in Eq.~(\ref{setintorho}), the system evolves according to classical stochastic theory from $\textbf{v}_{\xi}(v_{1},v_{2},...,v_{k},...,v_{d^2-1})$ to $\rho_{\mu}$. For a given initial state of a specific property, the final state can be written as the same form as Eqs.~(\ref{f3}) and (\ref{f12}) by $\rho_{\text{final}|v'_{k}}=\sum_{\mu}dP(\textbf{v}_{\mu})\rho_{\mu}$. Furthermore, the classical process makes $d^2-1$ complementary observables (e.g., the generalized Pauli matrices \cite{generalizedmatrices}) chosen for PT decomposable in terms of the final states $\rho_{\text{final}|v'_{k}}$. These observables then can be used to determine $\chi_C$ of the classical process for the $d$-level system.

\subsection*{Quantifying quantum-mechanical processes}

We now turn to the question of how to quantitatively characterize quantum-mechanical processes. Suppose that a process of interest is created and that its normalized process matrix, $\chi_{\text{expt}}$, is derived from experimentally available data using the PT procedure. If the experimental result cannot be described at all by any classical processes, then we say that $\chi_{\text{expt}}$ is a genuinely quantum process, denoted by $\chi_{Q}$ (Fig.~\ref{basicidea}a). To place this concept into a wider context, we introduce four different approaches for the quantitative characterization of $\chi_{Q}$ in $\chi_{\text{expt}}$:\\

\noindent (A1) Quantum composition:
\begin{equation}
\chi_{\text{expt}}=\alpha \chi_{Q}+(1-\alpha)\chi_{C},\label{composition}
\end{equation}
where $\alpha$ denotes the minimum amount of $\chi_{Q}$ that can be found in $\chi_{\text{expt}}$. The minimum amount of $\chi_{Q}$ that can be found in $\chi_{\text{expt}}$ is obtained by minimizing the following quantity via semi-definite programming (SDP) with MATLAB \cite{Lofberg, sdpsolver}:
\begin{equation}
\alpha\equiv \min_{\tilde{\chi}_{C}}\hspace{2pt}[1-\text{tr}(\tilde{\chi}_{C})],\label{alpha}
\end{equation}
such that
\begin{equation}
\chi_{\text{expt}}-\tilde{\chi}_{C}=\tilde{\chi}_{Q} \geq 0,\hspace{3pt}\rho_{\mu} \geq 0\hspace{0.5cm}\forall\mu,
\end{equation}
where $\tilde{\chi}_{Q}$ and $\tilde{\chi}_{C}$ are both unnormalized process matrices.\\

\noindent (A2) Process robustness:
\begin{equation}
\frac{\chi_{\text{expt}}+\beta\chi'}{1+\beta}=\chi_{C},\label{robustness}
\end{equation}
where $\beta$ represents the minimum amount of the noise process $\chi'$. The minimum amount of noise process is determined via SDP:
\begin{equation}
\beta\equiv \min_{\tilde{\chi}_{C}}\hspace{2pt}\left[\text{tr}(\tilde{\chi}_{C})-1\right],\label{beta}
\end{equation}
such that
\begin{equation}
\text{tr}(\tilde{\chi}_{C}) \geq 1,\hspace{3pt}\tilde{\chi}_{C}-\chi_{\text{expt}} \geq 0,\hspace{3pt}\rho_{\mu} \geq 0\hspace{0.5cm}\forall\mu.\label{betacondition}
\end{equation}
The first criterion in (\ref{betacondition}) guarantees that $\beta\geq0$, and the rest ensures that the noise $\chi'$ and the output states $\rho_{\mu}$ are positive semi-definite. For example, when $\tilde{\chi}_{C}-\chi_{\text{expt}}=0$, $\chi_{\text{expt}}$ is a genuinely classical process with $\beta=0$.\\

\begin{figure*}
\includegraphics[width=17.8cm]{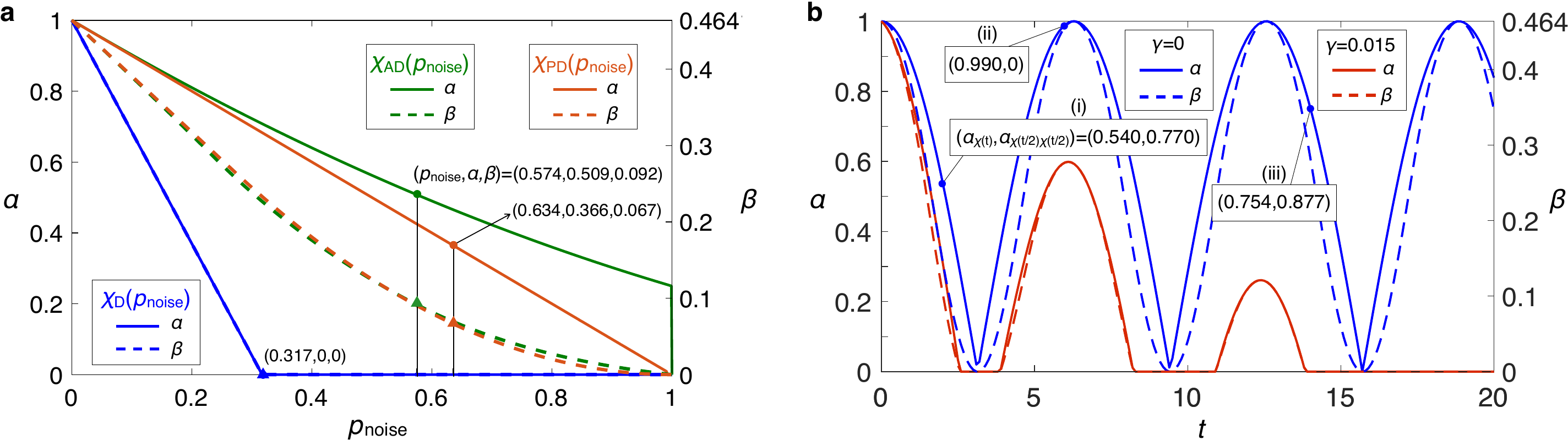}
\caption{Open quantum systems. \textbf{a} Quantum noise on a state-preserving quantum process. For zero noise intensity $p_{\text{noise}}=0$, $\chi_{\text{D}}$ (depolarization), $\chi_{\text{AD}}$ (amplitude damping), and $\chi_{\text{PD}}$ (phase damping) are identified as genuinely quantum, as an identity unitary transformation. $\alpha$ and $\beta$ for all the noise processes monotonically decrease with an increase in the noise intensity $p_{\text{noise}}$. These noise processes are identified as reliably close to the target state-preserving process if their $\alpha$ and $\beta$ are greater than certain thresholds as marked with $\bullet$ and $\blacktriangle$, respectively. See the property (P2). \textbf{b} Non-Markovian dynamics. Since $\alpha$ and $\beta$ monotonically decrease with time for Markovian dynamics, the non-Markovianity of $\chi_{\text{expt}}$ can be measured by integrating the positive derivative of $\alpha$ or $\beta$ with respect to time: $h_{q}(\Delta t)\equiv\int_{0;\dot{q}>0}^{\Delta t}\dot{q}dt$, for $q=\alpha,\beta$. As shown in Fig.~\ref{basicidea}b, we consider a system that is coupled to an environment with a state $p\left|0\right\rangle\left\langle 0\right|+(1-p)\left|1\right\rangle\left\langle 1\right|$ via a controlled-$Z$-like interaction $H=1/2\sum_{i,j=0}^{1}(-1)^{i\cdot j}\left|ij\right\rangle\left\langle ij\right|$ and depolarized with a rate $\gamma$. For example, we have $h_{\alpha}(15)\sim 0.86$ for $p=0.5$ and $\gamma=0.015$. (i)-(iii) illustrate the invalidation of $\alpha_{\chi_{\text{expt}}}=\alpha_{\chi_{2}\chi_{1}}$. Such detection is more sensitive than the existing non-Markovianity quantifiers, such as the  Breuer-Laine-Piilo (BLP) \cite{Breuer09} and Rivas-Huelga-Plenio (RHP) \cite{Rivas10} measures. For example, for $\gamma=0.25$ and $p=0.1$, we find that $\alpha_{\chi_{\text{expt}}}\neq\alpha_{\chi_{2}\chi_{1}}$ when $t<1.1$, whereas they certify the dynamics as Markovian. The certifications by the BLP and RHP measures are detailed in Ref. \cite{Chen15}. Indeed, our method is finer than the BLP and RHP measures for all the settings of $\gamma$ and $p$ considered therein.}
\label{noiseprocess}
\end{figure*}

\begin{figure}[h]
\includegraphics[width=8.9cm]{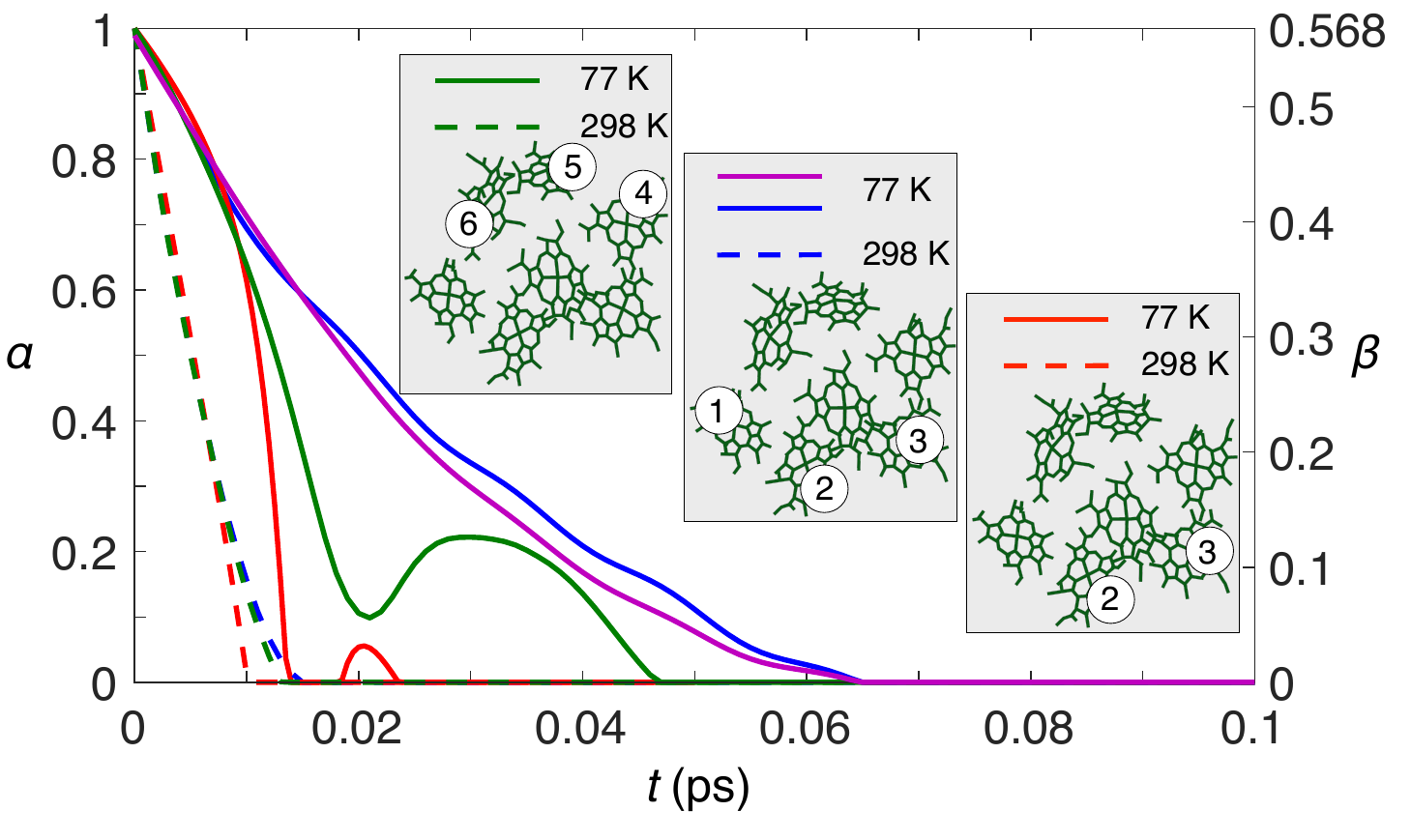}
\caption{Quantum transport in the FMO complex. We take two-site and three-site subsystems for examples and show how the amount of quantum transport ($\alpha$: green, blue, and red; $\beta$: purple) at temperatures of $77$ K (solid) and $298$ K (dash) varies with time ($t$) therein. A Lindblad master equation is used to model the dynamics of subsystem expressed in the site basis \cite{Plenio08}, including the coherent evolution, the dissipative recombination of exciton ($\chi_{\text{AD}}$) with a rate $\sim5\times 10^{-4}$ ps$^{-1}$ for all the sites, the dephasing interaction with the environment ($\chi_{\text{PD}}$), and the trapping of exciton in the reaction centre through site $3$ with a rate $6$ ps$^{-1}$. See Methods. The dephasing rates $2.1$ ps$^{-1}$ and $9.1$ ps$^{-1}$ corresponding to $77$ K and $298$ K, respectively, are considered.}
\label{fmo}
\end{figure}

\noindent (A3) Fidelity: $\chi_{\text{expt}}$ is detected as nonclassical close to a target quantum process $\chi_{Q_{T}}$ if the process fidelity satisfies that
\begin{equation}
F_{\text{expt}}\equiv \text{tr}(\chi_{\text{expt}}\chi_{Q_{T}})>F_{C}\equiv \max_{\chi_{C}} \text{tr}(\chi_{C}\chi_{Q_{T}}),\label{fidelity}
\end{equation}
or $\bar{F}_{s,\text{expt}}>\bar{F}_{s,C}$, stated in terms of the average state fidelity \cite{Gilchrist05} $\bar{F}_{s,\text{expt}(C)}= (dF_{\text{expt}(C)}+1)/(d+1)$. The concept of (\ref{fidelity}) is to rule out the best classical mimicry to an ideal target quantum process $\chi_{Q_{T}}$. Such capability of classical mimicry is evaluated by performing the following maximization task via SDP:
\begin{equation}
F_C\equiv \max_{\tilde{\chi}_C}\hspace{3pt}\text{tr}(\tilde{\chi}_C\chi_{Q_T}),
\end{equation}
such that
\begin{equation}
\text{tr}(\tilde{\chi}_{C})=1,\hspace{3pt}\rho_{\mu} \geq 0\hspace{0.5cm}\forall\mu.\label{fidelitycontraint}
\end{equation}
The first constraint in (\ref{fidelitycontraint}) ensures that $\tilde{\chi}_{C}$ is a normalized process matrix satisfying the definitions of the fidelity and a density operator.\\

\noindent (A4) Entropy: The process is certified as truly quantum mechanical if
\begin{equation}
S(\chi_{\text{expt}})<S_{C} \equiv\min_{\chi_{C}} S(\chi_{C}),
\end{equation}
where $S(\chi_{\text{expt}(C)})\equiv-\text{tr}(\chi_{\text{expt}(C)}\log\chi_{\text{expt}(C)})$. Here, $S_{C}$ can be analytically derived from the basic properties of $\chi_{C}$ and $S$. See Methods for the proof.\\

While the approaches (A1)-(A3) are based on different concepts and points of view, the following three properties of quantum and classical processes reveal close connections between the quantities derived from them:\\

\noindent (P1) If an experimental process consists of two sub-processes: $\chi_{\text{expt}}=\chi_{2}\chi_{1}$, where $\chi_{k}=\alpha_{k}\chi_{Qk}+(1-\alpha_{k})\chi_{Ck}$ for $k=1,2$, then both its $\alpha\equiv\alpha_{\chi_{2}\chi_{1}}$ and $\beta\equiv\beta_{\chi_{2}\chi_{1}}$ are not larger than that of the individual sub-process $\chi_{k}$, i.e., $\alpha_{\chi_{2}\chi_{1}}\leq\alpha_{k}$ and $\beta_{\chi_{2}\chi_{1}}\leq\beta_{k}$.\\

\noindent (P2) Not only does (A3) indicate whether $\chi_{\text{expt}}$ is reliable with respect to $\chi_{Q_{T}}$, both $\alpha$ and $\beta$ reflect the reliability of $\chi_{\text{expt}}$ as well: $\chi_{\text{expt}}$ is verified as reliable close to $\chi_{Q_{T}}$ if $\alpha>(F_{C}-F_{c})/(F_{q}-F_{c})$, where $F_{c}\equiv\text{tr}(\chi_{C}\chi_{Q_{T}})$ and $F_{q}\equiv\text{tr}(\chi_{Q}\chi_{Q_{T}})$. Similarly, if $\beta>(F_{C}-F_{c})/(F_{c}-F')$, where $F'\equiv\text{tr}(\chi'\chi_{Q_{T}})$, then $\chi_{\text{expt}}$ is identified as nonclassical and close enough to $\chi_{Q_{T}}$.\\

\noindent (P3) Suppose that, given a process $\chi_{\text{expt}}$, a classical process $\chi_{C}$ is derived from the definition of process robustness (A2). Only when $\chi_{\text{expt}}=\chi_{Q_{T}}$ the fidelity of $\chi_{C}$ and the target $\chi_{Q_{T}}$ may reach the classical upper bound of the process fidelity $F_{C}$.\\

The proof of (P1) is based on a condition under which two sub-processes can constitute a quantum-mechanical process. Since a classical process matrix is composed of the matrix elements specified by the classical theories for the initial system and the subsequent state transitions, the product of such matrix and any second process matrix is still a classical one since its matrix elements are linear combinations of those matrix elements of the classical process matrix.

A product of two sub-process matrices is quantum only when both the sub-processes are nonclassical. This implies that only the term $\chi_{Q2}\chi_{Q1}$ with an intensity $\alpha_{2}\alpha_{1}$ in the expansion of $\chi_{2}\chi_{1}$ is not a classical process. Then $\alpha_{\chi_{2}\chi_{1}}$ determined by SDP can be smaller than or equal to $\alpha_{2}\alpha_{1}$, which concludes that $\alpha_{\chi_{2}\chi_{1}}\leq\alpha_{k}$ for $k=1,2$. Compared with the individual sub-process $\chi_{k}$, the process robustness of $\chi_{\text{expt}}$ remains or decreases in its intensity $\beta_{\chi_{2}\chi_{1}}$ in response to such a non-increasing quantum composition.

For (P2), the conditions for a reliable process can be shown by using the reliability criterion, $F_{\text{expt}}>F_{C}$, and the basic definitions of $\alpha$, $\beta$ and the process fidelity. These conditions on $\alpha$ and $\beta$ can be represented in terms of average state fidelities as well.

The last property reminds us that, from the point of view of process robustness, only when $\chi_{\text{expt}}=\chi_{Q_{T}}$ the derived $\chi_{C}$ is the classical process that has the minimum deviation from $\chi_{Q_{T}}$ by an amount of noise $\beta$. Therefore the fidelity of $\chi_{C}$ and $\chi_{Q_{T}}$ may be larger than any other classical similarities and then equal to $F_{C}$.

\subsection*{Examples and applications}

The quantum operations formalism underlying our framework is a general tool for describing the dynamics experienced by either closed or open quantum systems in a wide variety of physical scenarios. Relying on this important feature, the utility and application of our formalism is illustrated by the following examples. The detailed derivations of these results are given in the Methods section.\\

\noindent (E1) Processes postulated in quantum mechanics. For any unitary transformation of finite size, we have $\alpha =1$ and $S=0$. By contrast, the projective measurements are identified as classical, i.e., $\alpha=0$. The POVM measurements \cite{Peres93} can be quantified in the same manner, depending on their concrete realizations.\\

\noindent (E2) Dynamics of open quantum systems and measuring non-Markovianity. (A1)-(A4) adapt naturally to unitary transformations affected by quantum noise in open quantum systems. See Fig.~\ref{noiseprocess}a. Moreover, $\alpha$ and $\beta$ provide a fruitful source of information regarding a measure and the finest detection of non-Markovian dynamics of a system coupled to an environment. If an experimental process is Markovian, $\chi_{\text{expt}}$ can be arbitrarily divided into sub-process matrices which satisfy the law of decomposition \cite{Breuer16,Vega17} $\chi_{\text{expt}}=\chi_{2}\chi_{1}$. For instance, the property (P1) implies that, with explicit reference to the passage of time for each sub-process, $\alpha$ and $\beta$ should monotonically decrease with time for a Markovian process. Hence, if we observe an increasing result from $\chi_{\text{expt}}$, then the process is verified as non-Markovian. Furthermore, there should be no differences between $\alpha$ and $\beta$ of the whole process, $\alpha_{\chi_{\text{expt}}}$ and $\beta_{\chi_{\text{expt}}}$, and those of the process composed of two sub-processes, $\alpha_{\chi_{2}\chi_{1}}$ and $\beta_{\chi_{2}\chi_{1}}$, for Markovian dynamics. An invalidation of this consistency reveals that the process is non-Markovian. See Fig.~\ref{noiseprocess}b.\\

\noindent (E3) Fusion of entangled photon pairs. Our framework inherits the far-reaching utility of the quantum operations formalism such that quantum dynamics can be explored by our novel quantification under a wide range of circumstances. The fusion of entangled photon pairs \cite{Pan12} superposes two individual photons in two different spatial modes at a polarizing beam splitter (PBS) and post-selects both outputs in different modes (Fig.~\ref{basicidea}d): $\alpha=1$, $\beta\sim0.657$, and $F_{C}\sim 0.604$.\\

\begin{table*}[t]
\centering
\label{IBMQC}
\begin{tabular}{|l|c|c|c|c|c|c|c|c|c|}
\hline
Methods & \multicolumn{7}{c|}{Single-qubit gate}                  & \multicolumn{2}{c|}{Two-qubit gate} \\ \cline{2-10}
                         & $U_{1}$ & $I$   & $X$   & $Y$   & $Z$   & $H$   & $T$   & $U_{2}$           & $\text{{\sc CNOT}}$            \\ \hline
$\alpha$                 & 1       & 0.884 & 0.941 & 0.871 & 0.863 & 0.836 & 0.799 & 1                 & 0.782           \\ \hline
$S$                      & 0       & 0.276 & 0.158 & 0.304 & 0.318 & 0.358 & 0.438 & 0                 & 1.302           \\ \hline
$F_{\text{expt}}$        & 1       & 0.959 & 0.980 & 0.960 & 0.953 & 0.947 & 0.934 & 1                 & 0.757           \\ \hline
\end{tabular}
\caption{Quantum gates in the quantum computer of IBM Q. We implement seven essential quantum gates with IBM Q. $U_{1}$ and $U_{2}$ represent the ideal (target) single-qubit and two-qubit gates, respectively. The process fidelities of all experimental cases considered here  \cite{Nielsen&Chuang00}: the identity gate ($I$), the Pauli operators ($X$, $Y$, $Z$), the Hadamard gate ($H$), the $\pi/8$ gate ($T$), and the $\text{{\sc CNOT}}$ gate, are all greater than the process fidelity thresholds $F_{C}=(1+\sqrt{3})/4\sim 0.683$ and $0.467$ (implying the average state-fidelity thresholds $\bar{F}_{s,C}\sim0.789$ and $0.574$, respectively), for single-qubit and two-qubit gate operations, respectively. Using (A4), conditioned on logarithms to base 2, their entropies are all less than the ultimate entropies of classical process $S_{C}=N$, where $N$ denotes the number of qubits being processed.}\label{table}
\end{table*}

\noindent (E4) Quantum transport in the Fenna-Matthews-Olson (FMO) complex. The FMO complex is a seven-site structure used by certain types of bacteria to transfer excitations from a light-harvesting antenna to a reaction centre (Fig~\ref{basicidea}c). Figure~\ref{fmo} suggests the first quantifications of nonclassical energy transfer in the FMO complex  \cite{Fenna75,Engel07}, where several pigments are chosen as a subsystem and single excitation transport is considered therein.\\

\noindent (E5) Quantum computation. We now examine concrete scenarios in which our formalism offers general benchmarks for quantum information. A valid quantum gate is specified by a unitary transformation ($\alpha=1$), and an arbitrary quantum gate can be expressed using single qubit and controlled-NOT ($\text{{\sc CNOT}}$) gates \cite{Nielsen&Chuang00} (Fig.~\ref{basicidea}e). We say that an experiment reliably implements quantum-information processing if $\chi_{\text{expt}}$ goes beyond the classical descriptions, such as superconducting circuits used for quantum information \cite{You05,Buluta11} and the quantum gates realized by the IBM quantum computer \cite{IBMQC}; see Table~\ref{table}.\\

\begin{figure}
\includegraphics[width=8.9cm]{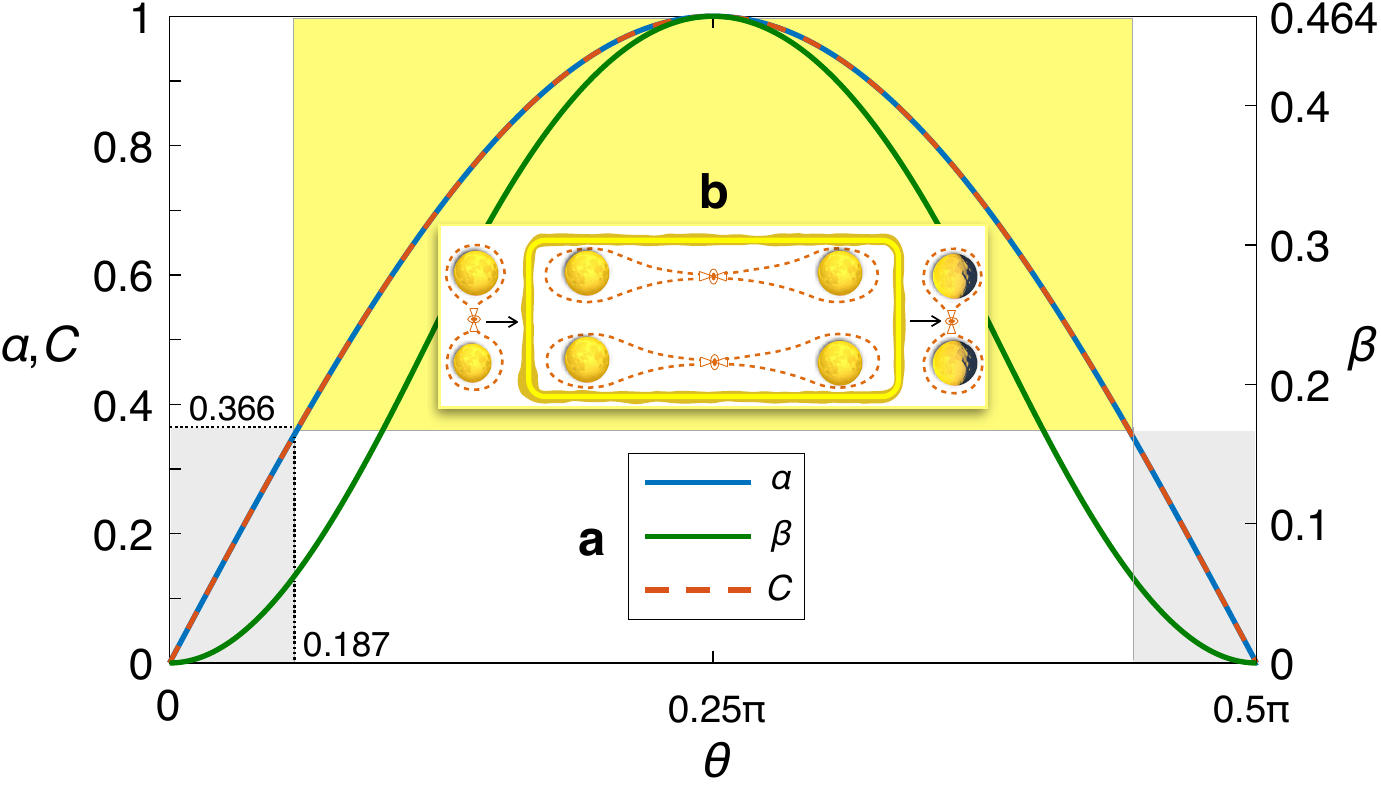}
\caption{Teleportation. \textbf{a} Without loss of generality we suppose a two-qubit system of the state $\left|\phi(\theta)\right\rangle=\cos\theta\left|00\right\rangle+\sin\theta\left|11\right\rangle$ is used for teleportation (Fig.~\ref{basicidea}f). The entanglement of $\left|\phi(\theta)\right\rangle$ measured by concurrence $C(\theta)=|\sin2\theta|$ can be strictly revealed by $\alpha$ and $\beta$ for the teleportation process. In particular, $\alpha$ exactly coincides with $C$. \textbf{b} Using the relation $C\geq 2F_{\text{expt}}-1$ \cite{Hofmann05}, as $C,\alpha > 2F_{C}-1\sim0.366$ (yellow region), two such entangled pairs enable teleportation of entanglement of qubits \cite{Lee00}. Compared to the steerable weight for quantifying EPR steering that are maximum for all pure entangled states \cite{Skrzypczyk14}, both $\alpha$ and $\beta$ can provide the qualities of entanglement previously shared between the sender and receiver for teleportation.}
\label{teleportation_noise}
\end{figure}

\noindent (E6) Quantum communication. An ideal qubit transmission between two parties acts as an identity unitary transformation on the transmitted qubit, which can be implemented by either sending qubits through an ideal communication channel \cite{BB84} or using teleportation \cite{Bennett96} (Fig.~\ref{basicidea}f) to move qubits around \cite{Gisin07}. For teleportation, both $\alpha$ and $\beta$ can reflect the qualities of entangled states shared between the sender and the receiver; see Fig.~\ref{teleportation_noise}a. In particular, our state-fidelity threshold is tighter than the well-known upper bound on the classical teleportation (i.e., $\bar{F}_{s,\text{expt}}=2/3\sim 0.667$ \cite{Massar95}) and guarantees faithful teleportation of the entangled qubits \cite{Lee00} (Fig.~\ref{teleportation_noise}b). Classical teleportation is a measure-prepare scenario in which the sender measures the unknown input state directly, and then sends the results to the receiver to prepare the output state \cite{Massar95,Pirandola15}. Such measure-prepare strategy attains its maximum process fidelity $F_{\text{expt}}=1/2$ at the output state fidelity $2/3$ for all arbitrary input states, and therefore is weaker than the best classical strategy with $F_{C}\sim 0.683$ and $\bar{F}_{s,C}\sim 0.789$ found by our method. Alternatively, the criterion $S(\chi_{\text{expt}})<1$ restricts the external disturbance to quantum-information processing, which remarkably coincides with the existing result for qubit transmission under coherent attacks \cite{Cerf02,Sheridan10,Chiu16}.

\subsubsection*{Usage and comparison}

As illustrated above, (A1)-(A4) can quantify the quantum nature of processes applied to a quantum systems in a wide variety of circumstances. The classification of an experimental process based upon its purpose  determines exactly which of the methods (A1-A4) is most useful. For example, compared to (A1) and (A2), for the task-oriented process aiming to experimentally realize quantum-information processing, (A3) can be used to directly evaluate whether $\chi_{\text{expt}}$ is close to $\chi_{Q_{T}}$ and superior to the best mimicry of a classical process. However, for general experiments with the purpose of investigating whether $\chi_{\text{expt}}$ is a quantum process, such as the energy transfer in FMO complex, (A1) and (A2) offer the advantage in performing two different types of quantitative analysis. The former focuses on the quantum composition of $\chi_{\text{expt}}$ and concretely determines the maximum proportion of the classical process of $\chi_{\text{expt}}$ in terms of $1-\alpha$. See Eqs.~(\ref{composition}) and (\ref{alpha}). Whereas, (A2) characterizes how close $\chi_{\text{expt}}$ is to a classical process in the sense that how large the minimum amount of noise, $\beta$, is required to make $\chi_{\text{expt}}$ classical [Eqs.~(\ref{robustness}) and (\ref{beta})]. Such a notion helps us understand and appreciate the roles $\alpha$ and $\beta$ have played in the quantitative analysis. For instance, it is easy to see why an experimental process may possess $\beta$ which is much smaller $\alpha$, as shown in Fig.~\ref{noiseprocess}a for $\chi_{\text{AD}}$ at $p_{\text{noise}}\rightarrow 1$.\\

\subsection*{Quantum correlations}

With our classical-process model (\ref{chic}) at hand, we can be precise regarding the statement about final states  generated by a generic classical process, and uncover new characteristics of quantum states. Let us consider a composite system of $N$ qubits and divide the system into two groups, $A$ and $B$, consisting of $n_{A}$ and $n_{B}$ qubits, respectively, where $n_{B}\geq 1$, and $n_{A}+n_{B}=N$. An $N$-qubit state is called $\chi_{C}$-nonclassical iff it cannot be generated by performing any classical processes on each qubit in $A$: $\rho_{\chi_{C}}^{(\kappa)}=\chi_{C}^{(A)}(\rho_{\text{initial}}^{(\kappa)})$, where $\chi_{C}^{(A)}$ denotes any operation composed of classical processes for each single qubit in $A$ on an initial state $\rho_{\text{initial}}^{(\kappa)}$ (Fig.~\ref{basicidea}g), and $\kappa$ signifies the bipartition type for $A$ and $B$. Otherwise, the state is called the $\chi_{C}$-classical state. When considering all the possible partitions of $\kappa$, we call a state genuinely multipartite $\chi_{C}$-nonclassical iff it cannot be represented by $\rho_{\chi_{C}}=\sum_{\kappa}p_{\kappa}\rho_{\chi_{C}}^{(\kappa)}$ for all possible bipartitions and probability distributions of $p_{\kappa}$. The basic concept behind $\chi_{C}$-classical states can be considered a hybrid of separable-states \cite{Werner89,Horodecki09} and the local hidden state (LHS) \cite{Wiseman07,Reid09,Cavalcanti16} models, implying a new property between genuine multipartite entanglement and genuine multipartite Einstein-Podolsky-Rosen (EPR) steering \cite{He13,Li15}, as shown in Methods.

A witness operator that detects genuinely multipartite $\chi_{C}$-nonclassical states that are close to a pure target state $\left|\psi_{T}\right\rangle$ is given by
\begin{equation}
\mathcal{W}=w_{\chi_{C}}\mathbbm{1}-\left|\psi_{T}\right\rangle\!\!\left\langle\psi_{T}\right|,\label{witness}
\end{equation}
where $\mathbbm{1}$ is the identity operator for $N$ qubits, and
\begin{equation}
w_{\chi_{C}}\equiv\max_{\rho_{\chi_{C}}}\left\langle\psi_{T}\right|\rho_{\chi_{C}}\left|\psi_{T}\right\rangle.\label{witnesskernel}
\end{equation}
Thus, any experimental state $\rho_{\text{expt}}$ with $\text{tr}\big(\mathcal{W}\rho_{\text{expt}}\big)<0$, i.e., the quality in terms of the state fidelity $F_{s,\text{expt}}>w_{\chi_{C}}$ is a truly multipartite $\chi_{C}$-nonclassical state close to $\left|\psi_{T}\right\rangle$. For example, we have $w_{\chi_{C}}=(1+\sqrt{3})/4\sim0.683$ for the Greenberger-Horne-Zeilinger (GHZ) states of three qubits \cite{GHZstates}. We show how to determine $w_{\chi_{C}}$ in Methods.

\subsection*{Characterizing quantum states with process quantifications}

Note that the characterization of quantum states can benefit by including a quantum-mechanical process. For example, EPR steering \cite{Wiseman07,Reid09,Cavalcanti16} can be enlarged by considering that the untrusted party proceeds to perform a quantum-information process, e.g., teleportation (Fig.~\ref{teleportation_noise}) or one-way quantum computing \cite{Briegel09}. Moreover, the model of quantum process explicitly sheds light on the temporal analogue of EPR steering \cite{Chen14,Li15b,Bartkiewicz16,Li16} and naturally provides its optimum quantification, which cannot be provided by existing methods \cite{Chen16}.

Let us take the temporal steering for single systems transmitted by a sender, Alice, to a receiver, Bob as an example. The concepts of EPR steering and the LHS model are used for timelike separations between Alice and Bob. For instance, in the temporal version of the LHS model, the joint probability of observing $v_{a}$ by Alice at time $t_{a}$ and $v_{b}$ by Bob at time $t_{b}$, where $t_{a}<t_{b}$, is specified by $P(v_{a,t_{a}},v_{b,t_{b}})=\sum_{\mu}p_{\mu}P(v_{a,t_{a}}|\mu)P(v_{b,t_{b}}|\sigma_{\mu})$, where $\sigma_{\mu}$ denotes the state of system held by Bob. It is easy to see that this representation of the joint probability can be described through Eq.~(\ref{resultingstate}) in the model of classical process, i.e.,  $P(v_{a,t_{a}},v_{b,t_{b}})=\sum_{\mu}P(v_{a,t_{a}})\Omega_{v_{a,t_{a}}\mu}P(v_{b,t_{b}}|\sigma_{\mu})$.

Our formalism can explain the rationale behind the temporal version of the LHS model and show the result that cannot be provided by existing methods, such as the optimal quantification of temporal steering. The approach introduced in \cite{Chen16} is parallel to the method for quantifying EPR steering. The state of Bob's system conditioned on Alice's result $v_{a,t_{a}}$ can be described by
\begin{equation}
\sigma^{T}_{v_{a,t_{a}}}=\tau\sigma^{T,S}_{v_{a,t_{a}}}+(1-\tau)\sigma^{T,US}_{v_{a,t_{a}}}.\nonumber
\end{equation}
Without loss of generality we may suppose $v_{a,t_{a}}=v_{k}$ for the state of the $k$th physical property of Alice's system. Each unnormalized unsteerable state in the unsteerable assemblage $\{\sigma_{v_{k}}^{T,US}\}$ can be written in the hidden-state form: $\sigma_{v_{k}}^{T,US}=\sum_{\mu}p_{\mu}P(v_{k}|\mu)\sigma_{\mu}$. See Eq. (3) in the work \cite{Chen16}. The temporal steerable weight $\tau$ measures the ``steerability in time'' for a given assemblage $\{\sigma^{T}_{v_{k}}\}$, and is obtained by an minimization procedure with respect to $\{\sigma_{v_{k}}^{T,S}\}$. Such approach to describing temporal steering in terms of $\tau$ is nonoptimal in the sense that it depends on the number and types of measurements being used for $v_{k}$.

Our method quantifies the optimal temporal steering. One can use $\alpha$ to represent the maximum temporal steering that can be found in a process through single systems. It is easy to see that, after a process $\chi_{\text{expt}}$ (\ref{composition}), an initial state $\rho_{\text{initial}}$ becomes
\begin{equation}
\rho_{\text{final}}=\alpha \chi_{Q}(\rho_{\text{initial}})+(1-\alpha)\chi_{C}(\rho_{\text{initial}}).\nonumber
\end{equation}
To faithfully show the effects of a process on the system, $\rho_{\text{final}}$ is assumed to be pure. Then $\chi_{Q}(\rho_{\text{initial}})$ is still pure to go beyond the description (\ref{resultingstate}). Whereas, by Eqs.~(\ref{resultingstate}) and (\ref{f3}), $\chi_{C}(\rho_{\text{initial}})$ follows the classical model, which explains the unsteerable state by
\begin{eqnarray}
\chi_{C}(\rho_{\text{initial}})&=&P(v'_{k})^{-1}\sigma_{v'_{k}}^{T,US}\nonumber\\
&=&\rho^{(c)}_{\text{final}|v'_{k}}=\sum_{\mu}\sum_{\xi}P(\textbf{v}_{\xi})P(v'_{k}|\textbf{v}_{\xi})\delta_{\xi\mu}\rho_{\mu}.\nonumber
\end{eqnarray}
Compared with the steerable weight, $\alpha$ is optimum for all input states and therefore larger than $\tau$ under a given assemblage $\{\sigma^{T}_{v_{k}}\}$ with finite elements. See Table~\ref{temporal} for concrete illustrations and comparison.

\begin{table}[t]
\centering
\begin{tabular}{|l|c|c|c|}
\hline
Approaches & \multicolumn{3}{c|}{Target channels} \\ \cline{2-4}
                            & $I$      & $H$      & $T$     \\ \hline
$\alpha$                    & 0.884    & 0.836    & 0.799   \\ \cline{2-4}
$\tau$$_{8}$                   & 0.769    & 0.688    & 0.603   \\ \cline{2-4}
$\tau$$_{7}$                   & 0.768    & 0.687    & 0.602   \\ \cline{2-4}
$\tau$$_{6}$                   & 0.767    & 0.677    & 0.602   \\ \cline{2-4}
$\tau$$_{5}$                   & 0.766    & 0.677    & 0.600   \\ \cline{2-4}
$\tau$$_{4}$                   & 0.765    & 0.674    & 0.599   \\ \cline{2-4}
$\tau$$_{3}$                   & 0.764    & 0.671    & 0.597   \\ \cline{2-4}
$\tau$$_{2}$                   & 0.688    & 0.522    & 0.391   \\ \hline
\end{tabular}\caption{Comparison of quantum composition ($\alpha$) and temporal steerable weight ($\tau$). We consider how temporal steering is demonstrated by single qubits undergoing practical channels in IBM Q \cite{IBMQC} and quantified in terms of $\alpha$ and $\tau$ \cite{Chen16}. These experimental channels are created with respect to the gates \textit{I}, \textit{H} and \textit{T}. Here, the subscript of $\tau_{n}$ indicates that an assemblage $\{\sigma^{T}_{v_{k}}\}$ with $n$ elements has been used to determine $\tau$ for the experimental states. In determining $\tau$$_{2}$, the eigenstates of \textit{X} and \textit{Y} were chosen as the input states of the experimental channels. For $n>2$, the eigenstates of \textit{X}, \textit{Y} and \textit{Z} together with $2n-6$ randomly-chosen pure states as the input states were used in calculating $\tau_{n}$. Since our framework is optimal, $\alpha$ is higher than $\tau_{n}$ for each experimental channel, irrespective of the type and number of input states chosen to obtain $\tau$.}\label{temporal}
\end{table}

\section*{Discussion}

In this work, we clarified and broadened basic ideas behind the distinction between classicality and quantumness, addressing the most basic problem of how to quantitatively characterize physical processes in the quantum world. We showed for the first time that quantum-mechanical processes can be quantified. We revealed that such quantification can have profound implications for the understanding of quantum mechanics, quantum dynamics, and quantum-information processing. Our approach is more general than many existing methods, and much broader in scope than theories based on state analysis. Our formalism is applicable in all physical processes described by the general theory of quantum operations, including but not limited to the fundamental processes postulated in quantum mechanics, the dynamics of open quantum systems, and the task-orientated processes for quantum technology. This far-reaching utility of our framework enables us to explore quantum dynamics under a wide range of circumstances, such as the fusion of entangled photon pairs and the energy transfer in a photosynthetic pigment-protein complex. In addition, our formalism enables quantum states to be characterized in new ways, to uncover new properties of both composite and single systems.

Since all of our approaches are experimentally feasible, they can be readily implemented in a wide variety of the present experiments \cite{Coherence08,Pan12}, such as the quantum channel simulator \cite{Lu17} and ground-to-satellite teleportation \cite{Ren17}. However, it is important to have a clear appreciation for the limitations of the quantum operations formalism underlying the constructions for our framework, such as the assumption of a system and environment initially in a product state\cite{Nielsen&Chuang00, Vega17, royer1996reduced}. Such prior knowledge about the system and environment is therefore required to perform process quantifications.

For future studies and applications of our concept and methods, we anticipate their use in general physical processes, such as superpositions \cite{Coherence08}, asymmetries \cite{Peres93}, and randomness \cite{Acin16}. Using modern machine learning techniques \cite{Carrasquilla17} could improve the performance and scalability of PT and quantification of complex system processes, such as those found in condensed-matter physics. Furthermore, provided the measurement outcomes are continuous and unbound, it is enlightening to attempt to extend our formalism to encompass the quantifications of nonclassical processes in harmonic systems such as nanomechanical resonators \cite{Johansson14}. These essential elements could promote novel recognition and classification of physical processes with a generic process quantifier.

\section*{Methods}
\subsection*{Fundamental processes in quantum mechanics and quantum noise}

The evolution of quantum systems and the application of quantum measurements are two essential kinds of processes prescribed by quantum mechanics. The evolution of a closed quantum system and the effects of measurements are described by a unitary transformation $U$ and a collection of measurement operators $\textbf{M}=\{M_{m}\}$, respectively, where the index $m$ denotes the measurement outcomes that obtained in the experiment \cite{Shankar94}. For any $U$ of finite size, its process matrix $\chi_{U}$ always can be expressed in an orthonormal basis as a diagonal matrix with only one non-vanished matrix element, i.e., $S(\chi_{U})=0$, which makes any classical process matrices unable to represent $\chi_{U}$ at all and implies that $\alpha =1$. When unitary transformations are affected by quantum noise to become noise processes, their quantification is dependent on the type of noise and the noise intensity, as shown in Fig.~\ref{noiseprocess}a. The three important examples of quantum noise considered therein: depolarization ($\chi_{\text{D}}$), amplitude damping ($\chi_{\text{AD}}$), and phase damping ($\chi_{\text{PD}}$), are defined as follows \cite{Nielsen&Chuang00}:
\begin{eqnarray}
\chi_{\text{D}}(\rho)\!&=\!&(1\!-\!\frac{3}{4}p_{\text{noise}})I\rho I+\frac{1}{4}p_{\text{noise}}(X\rho X+Y\rho Y+Z\rho Z), \nonumber\\
\chi_{\text{PD}}(\rho)\!&=\!&(1\!-\!\frac{1}{2}p_{\text{noise}})I\rho I+\frac{1}{2}p_{\text{noise}}Z\rho Z,\\
\chi_{\text{AD}}(\rho)\!&=\!&K\rho K^{\dag}\!+\frac{p_{\text{noise}}}{4}(X\!+\!iY)\rho (X\!-\!iY),\nonumber
\end{eqnarray}
where $K=(\frac{1+\!\sqrt{1\!-\!p_{\text{noise}}}}{2}I\!+\!\frac{1-\!\sqrt{1\!-\!p_{\text{noise}}}}{2}Z)$.

Projective measurements is an important special case of the measurement postulate where the measurement operators satisfy the conditions of projectors, $M_{m}M_{m'}=\delta_{mm'}M_{m}$ and $M_{m}^\dag=M_{m}$. Since the process matrix $\chi_{M_{m}}$ of a given $M_{m}$ expressed in $\textbf{M}$ is diagonal, this matrix can be fully described by a classical process matrix $\chi_{C}$. Thus the process of the state changes effected by the projector $M_{m}$ is identified as classical, i.e., $\alpha=0$. On the other hand, the quantification of the positive operator-valued measure (POVM) measurements depends on the realization or structure of $M_{m}$ under consideration.

\subsection*{Fusion of entangled photon pairs}

The fusion of entangled photon pairs combines quantum interference with post selection for photon pairs to provide an excellent experimental method for fusing different entangled pairs as genuinely multipartite entangled photons of multi-photon Greenberger-Horne-Zeilinger (GHZ) states (illustrated in Fig.~\ref{basicidea}d) \cite{Lu07}. When superposing two individual photons in two different spatial modes at a polarizing beam splitter (PBS) that transmits $H$ (horizontal) and reflects $V$ (vertical) polarization, a coincidence detection of the both outputs in different modes implements a photon fusion described by $M_{\text{PF}}\equiv M_{H1}\otimes M_{H2}+M_{V1}\otimes M_{V2}$, where $M_{mk}=\left|m\right\rangle_{\!kk}\!\!\left\langle m\right|$ for $m=H,V$ and $k=1,2$. It is nonclassical: $\alpha=1$, $\beta\sim0.657$, and $F_{C}\sim 0.604$. The photonic Bell-state and GHZ-state analyzing processes \cite{Pan12} can be quantified by the same method. The Bell-state analyzer, which exploits quantum interference due to the bosonic nature of photons at a 50:50 beam splitter, has the same results as the photon fusion. As an extended process of photon fusion, the basic process underlying the GHZ-state analyzer can be described by $M_{N\text{GHZ}}\equiv \bigotimes_{k=1}^{N}M_{Hk}+\bigotimes_{k=1}^{N}M_{Vk}$ for $N$-photon GHZ states. For instance, it is identified as a truly nonclassical process with $\alpha=1$, $\beta\sim 0.798$, and $F_{C}\sim0.556$ for $N=3$.

\subsection*{Quantum transport in the FMO complex}

Distinguishing quantum from classical processes for the energy transport in the FMO pigment-protein complex \cite{Fenna75,Engel07} is crucial to appreciate the role the nonclassical features play in biological functions \cite{Lambert13}. Figure~\ref{fmo} shows that the quantum transport in the FMO complex is identified and quantified on considerable timescales. Here we assume that the FMO system is in the single-excitation state of the form \cite{Plenio08,Caruso09}:
\begin{eqnarray}
\rho=\sum_{i,j\in\{1,...,7,E\}}\rho_{i,j}\left|i\right\rangle\!\!\left\langle j\right|,
\end{eqnarray}
where $\left|j\right\rangle$ in the site basis $\{ \left|i\right\rangle\}^{7}_{i=1}$ represents the excitation is shown at site $j$ and $\left|E\right\rangle$ means an empty state in the absence of excitation. The time evolution of the state $\rho$ is described by the Lindblad master equation:
\begin{eqnarray}
\dot{\rho}=-\text{i}[H,\rho]+\mathcal{L}_\text{diss}(\rho)+\mathcal{L}_\text{sink}(\rho)+\mathcal{L}_\text{deph}(\rho).\label{FMOevolution}
\end{eqnarray}
The Hamiltonian \textit{H} for the coherent transfer of single excitation between sites is \cite{Adolphs06}
 \begin{equation}
H\equiv
\left [ \begin{matrix}
                 215 & -104.1 & 5.1 & -4.3 & 4.7 & -15.1 & -7.8\\
                 -104.1 & 220 & 32.6 & 7.1 & 5.4 & 8.3 & 0.8\\
                 5.1 & 32.6 & 0 & -46.8 & 1.0 & -8.1 & 5.1\\
                 -4.3 & 7.1 & -46.8 & 125 & -70.7 & -14.7 & -61.5\\
                 4.7 & 5.4 & 1.0 & -70.7 & 450 & 89.7 & -2.5\\
                 -15.1 & 8.3 & -8.1 & -14.7 & 89.7 & 330 & 32.7\\
                 -7.8 & 0.8 & 5.1 & -61.5 & -2.5 & 32.7 & 280
               \end{matrix} \right].
\end{equation}
The incoherent dynamics is described by the three Lindblad superoperators $\mathcal{L}_\text{diss}$, $\mathcal{L}_\text{sink}$, and $\mathcal{L}_\text{deph}$ in (\ref{FMOevolution}). The superoperator $\mathcal{L}_{\text{diss}}$ specifies the dissipative recombination of excitation by
\begin{eqnarray}
\mathcal{L}_\text{diss}(\rho)=\sum_{i=1}^{7}\Gamma_i(2\left|E\right\rangle\!\!\left\langle i\right|\rho\left|i\right\rangle\!\!\left\langle E\right|-\{\left|i\right\rangle\!\!\left\langle i\right|,\rho\}),
\end{eqnarray}
where the recombination rate at each site is $\Gamma_i\sim5\times10^{-4}\ \text{ps}^{-1}$\cite{Caruso09}. The second Lindblad superoperator describes the trapping of excitation from site 3 to the reaction centre:
\begin{eqnarray}
\mathcal{L}_\text{sink}(\rho)=\Gamma_\text{sink}(2\left|E\right\rangle\!\!\left\langle 3\right|\rho\left|3\right\rangle\!\!\left\langle E\right|-\{\left|3\right\rangle\!\!\left\langle 3\right|,\rho\}),
\end{eqnarray}
with the trapping rate $\Gamma_\text{sink}\sim6\ \text{ps}^{-1}$ \cite{Caruso09}. The superoperator $\mathcal{L}_\text{deph}$ shows the dephasing interaction with the environment by
\begin{eqnarray}
\mathcal{L}_\text{deph}(\rho)=\sum_{i=1}^{7}\gamma_i(2\left|i\right\rangle\!\!\left\langle i\right|\rho\left|i\right\rangle\!\!\left\langle i\right|-\{\left|i\right\rangle\!\!\left\langle i\right|,\rho\}),
\end{eqnarray}
where the dephasing rates at each site are $\gamma_i\sim2.1\ \text{ps}^{-1}$ \text{and} $9.1\ \text{ps}^{-1}$ for 77 K and 298 K, respectively \cite{Rebentrost09,FMO}.

To quantify the quantum transfer in the FMO system, taking the subsystem composed of the pigments 4, 5 and 6 for example, we implement PT on this subsystem to get the corresponding process matrix $\chi_{\text{expt}}(t)$. We first use eight properties which correspond to a set of eight complementary observables $\{V_{k}\}$ where each one has three possible outcomes $v_{k}\in\{+1,0,-1\}$ to describe such a three-dimensional subsystem. As illustrated at the beginning of the Methods section, a process matrix $\tilde{\chi}_{\text{expt}}(t)$ can be obtained by analyzing the outputs of the eight complementary observables from the process: $V_{k}\rightarrow V_{k,\text{expt}}(t)\equiv\rho^{(\text{expt})}_{\text{final}|v'_{k}=+1}(t)-\rho^{(\text{expt})}_{\text{final}|v'_{k}=-1}(t)$, where $\rho^{(\text{expt})}_{\text{final}|v'_{k}}(t)$ denotes the eigenstate of $V_{k}$ corresponding to the eigenvalue $v'_{k}$ under the time evolution specified by Eq.~(\ref{FMOevolution}). It is clear that $V_{k,\text{expt}}(0)=V_{k}$. Note that, since the excitation can transfer between all the seven pigments and eventually leave the subsystem, the process matrix $\tilde{\chi}_{\text{expt}}(t)$ derived from $V_{k,\text{expt}}(t)$ is not trace-preserving. The trace of $\tilde{\chi}_{\text{expt}}(t)$ specifies a probability of observing single excitation transport in the subsystem \cite{Bongioanni10}. Here our approaches (A1) and (A2) are applied to quantify the normalized process matrix $\chi_{\text{expt}}(t)=\tilde{\chi}_{\text{expt}}(t)/ \text{tr}(\tilde{\chi}_{\text{expt}}(t))$ under time evolution, as shown in Fig.~\ref{fmo}. With our tool at hand, one can quantitatively investigate how the characteristics of the FMO system change under a variety of external operations or noise processes \cite{Chen13,Mourokh15,Chen17}.

\subsection*{Criterion for reliable qubit transmission}

For the threshold $S_{C}= 1$ for single two-level systems ($d=2$), the classical processes with the minimum entropy $S_{C}$ show that the maximum mutual dependence between the sender and receiver's results of two complementary measurements: $I_{SR}\equiv\sum_
{k=1}^{2}I_{S_{k}R_{k}}$, is restricted by $I_{SR,C}=1$, where $I_{S_{k}R_{k}}$ denotes the mutual information between their results of the $k$th measurement. Hence $I_{SR}>I_{SR,C}$ indicates that their communication process is reliable. For example, considering a phase damping channel $\chi_{\text{PD}}$ with noise intensity $p_{\text{noise}}=1$ which is identified as a classical process, we have the mutual information $I_{S_{1}R_{1},C}=1$ measured in the basis $\{\left|0\right\rangle,\left|1\right\rangle\}$ and $I_{S_{2}R_{2},C}=0$ in the basis $\{\left|+\right\rangle,\left|-\right\rangle\}$ where $\left|\pm\right\rangle=(\left|0\right\rangle\pm\left|1\right\rangle)/\sqrt{2}$.
When rephrasing $I_{SR}$ in terms of the average state fidelity $F_{s}$ and the error rate $D=1-F_{s}$ by
\begin{eqnarray}
I_{S_{1}R_{1},C}+I_{S_{2}R_{2},C}=2(F_{s}\text{log}_2F_{s}+D\text{log}_2D),
\end{eqnarray}
the classical threshold $I_{SR,C}=1$ provides an upper bound of the error rate for reliable communication as $D=0.110$. Importantly, this criterion coincides with the existing result for quantum communications under coherent attacks \cite{Cerf02,Sheridan10,Chiu16}.

\subsection*{Comparison of entanglement, steering and $\chi_{C}$-nonclassical correlations}

We first assume that the measurement outcomes for each qubit correspond to some observable with a set of eigenvalues $\{v_{a,k}\}$ or $\{v_{b,k}\}$ for the $k$th qubit in $A$ and $B$, respectively. The classical realistic elements $\textbf{v}_{\xi}$ and $\Omega_{\textbf{v}_{\xi}\mu}$ in a classical process performed on the $k$th qubit in $A$ prescribe the initial state of qubit with $v_{a,k}=v'_{a,k}$ a final state composed of states $\rho_{\mu}$, as shown in Eqs. (\ref{resultingstate}) and (\ref{chic}) in the main text. After $\chi_{C}^{(A)}$ has been done on $\rho_{\text{initial}}^{(\kappa)}$, the corresponding characteristics of states for $A$ and $B$ jointly can be revealed by considering the joint probabilities of obtaining outcomes of the measurements $v_{A}=\{v_{a,k}|k\in \text{n}_{A}\}$ and $v_{B}=\{v_{b,k}|k\in \text{n}_{B}\}$:
\begin{equation}
P(v_{A},v_{B})=P\big(v_{A},v_{B}|\{\sum_{\mu}\Omega_{v'_{a,k}\mu}\rho_{\mu}|k\in \text{n}_{A}\},\rho_{\text{initial}}^{(\kappa)}\big),\label{chicmodel}
\end{equation}
where $\text{n}_{A}=\{1,2,...,n_{A}\}$ and $\text{n}_{B}=\{1,2,...,n_{B}\}$.

The nonseparability of quantum states (sometimes called entanglement) \cite{Werner89,Horodecki09} and the EPR steering \cite{Wiseman07,Reid09,Cavalcanti16} go beyond the predictions of the model of separable states and the local hidden state (LHS) model \cite{Wiseman07}, respectively. The basic concept behind Eq.~(\ref{chicmodel}) can be considered a hybrid of these models. Without loss of generality, we consider the case for two particles ($N=2$). Compared to the states of particle $A$ that are determined by shared variables $\mu$ such that $P(v_{a},v_{b})=\sum_{\mu}p_{\mu}P(v_{a}|\mu)P(v_{b}|\sigma_{\mu})$ holds in the LHS model, the output states of $\chi_{C}$ involving $\rho_{\mu}$ are described by density matrices according to the prescribed realistic elements $\Omega_{v_{a,k}\mu}$ in the $\chi_{C}$-nonclassical model; see Eq.~(\ref{chicmodel}). While these states in the $\chi_{C}$-nonclassical model and those in the separable-state model which predicts that $P(v_{a},v_{b})=\sum_{\mu}p_{\mu}P(v_{a}|\rho_{\mu})P(v_{b}|\sigma_{\mu})$, are represented by density operators, $A$ and $B$ do share $\mu$ in the latter but $A$ and $B$ do not in the former. For these differences, the $\chi_{C}$-nonclassical correlation is stronger than nonseparability, but EPR steerability can be stronger than or equal to the $\chi_{C}$-nonclassical correlation. Here we illustrate such hierarchy by showing concrete quantum states of multipartite systems with the witness operators $\mathcal{W}$. In Eqs.~(\ref{witness}) and (\ref{witnesskernel}), the maximum similarity between $\left|\psi_{T}\right\rangle$ and $\rho_{\chi_{C}}$ can be explicitly determined by
\begin{equation}
w_{\chi_{C}}=\max_{\rho_{\chi_{C}}^{(\kappa)}}\left\langle\psi_{T}\right|\rho_{\chi_{C}}^{(\kappa)}\left|\psi_{T}\right\rangle,\label{witnesskernel2}
\end{equation}
which is equivalent to finding the best operational strategy for $A$ and $B$ such that a target state after the action on $A$ is closest to the original. As $n_{A}=1$ (i.e., $n_{B}=N-1$), $w_{\chi_{C}}$ is obtained by evaluating the maximum overlap $w_{\chi_{C}}=\max_{\kappa,\chi_{C}}\left\langle\psi_{T}\right|\chi_{C}^{(A)}(\left|\psi_{T}\right\rangle\!\!\left\langle\psi_{T}\right|)\left|\psi_{T}\right\rangle$ through SDP. For the three-qubit GHZ states \cite{GHZstates}, we have $w_{\chi_{C}}\sim0.683$ which is grater than the maximum value that can be attained for biseparable states $1/2$ \cite{Guhne09} and equal to the threshold for genuinely multipartite EPR steering \cite{Li15}. When taking $W$ states as the target state, whereas the identified EPR steerability is stronger than the $\chi_{C}$-nonclassical correlation. For example, $w_{\chi_{C}}\sim0.717$ for $N=3$ is grater than the threshold of $2/3$ for genuine tripartite entanglement \cite{Guhne09} but is weaker than the upper bound of $(1+\sqrt{2})/3\sim0.805$ that can be attained by non-genuine tripartite EPR steering \cite{Li15}.

\section*{Acknowledgements}

We are grateful to S.-L. Chen, Y.-N. Chen, C.-H. Chou, S.-Y. Lin, H. Lu, H.-S. Goan, O. G\"uhne, L. Neill and F. Nori for helpful comments. This work is partially supported by the Ministry of Science and Technology, Taiwan, under Grant Numbers MOST 104-2112-M-006-016-MY3.

\end{document}